# How Big Are Peoples' Computer Files? File Size Distributions Among User-managed Collections

**Jesse David Dinneen**
Humboldt-Universität zu Berlin, Germany
jesse.dinneen@hu-berlin.de

**Ba Xuan Nguyen**
Posts and Telecommunications Institute of Technology, Vietnam
nguyenxuanba@ptithcm.edu.vn

## ABSTRACT

Improving file management interfaces and optimising system performance requires current data about users' digital collections and particularly about the *file size distributions* of such collections. However, prior works have examined only the sizes of *system* files and users' *work* files in varied contexts, and there has been no such study since 2013; it therefore remains unclear how *today's* file sizes are distributed, particularly *personal* files, and further if distributions differ among the major operating systems or common occupations. Here we examine such differences among 49 million files in 348 user collections. We find that the average file size has grown more than ten-fold since the mid-2000s, though most files are still under 8 MB, and that there are demographic and technological influences in the size distributions. We discuss the implications for user interfaces, system optimisation, and PIM research.

## INTRODUCTION

An ongoing concern in personal information management (PIM) research is people's digital file collections, especially how and why people manage them, what the resulting collections are like, and how therefore to best support people with improved interfaces and services (Jones *et al.*, 2017). For example, studies of people's file collections have looked at their size, how the folders are organised, and the importance people attribute to them (Bergman *et al.*, 2010; Cushing, 2013; Dinneen *et al.*, 2019). Even the *sizes* of people's files are interesting, as they tell us about the nature of the items users manage and retrieve everyday, including how those items occupy storage space and grow over time and what simulated collections (used in testing new management tools) should look like if they are to be ecologically valid (Chernov *et al.*, 2008; Dinneen & Julien, 2020). More particularly, understanding collections' *composition* in terms of different file sizes – also known as the *file size distribution* (FSD) – can be used to improve the features and performance of file managers, desktop search algorithms, file recommender systems, file backup software, etc. We provide more detailed examples and concrete suggestions when discussing our results.

For the use of FSD data to be effective in such cases, the data must reflect the kinds of files the software will be used with, namely, *files in users' collections*. However, the vast majority of studies of FSD, conducted with the goal of optimising system performance (e.g., by choosing a block size according to the most frequent file sizes), have typically examined files that are atypical for most users, like system and server files, simulated network files, or files in high-performance computing environments (Downey, 2001; Evans & Kuenning, 2002; Gribble *et al.*, 1998; Harter *et al.*, 2011; Mitzenmacher, 2004; Ousterhout *et al.*, 1985; Roselli *et al.*, 2000; Satyanarayanan, 1981; Smith & Seltzer, 1981; Vogels, 1999; Welch & Noer, 2013). Very few studies have examined the FSD among *users' files*, and among those, all examine only users' *work* collections (which can contain different contents than collections used for personal matters; Dinneen & Julien, 2019). Further, past works' population samples vary greatly, precluding a comparison of results from different kinds of collections (e.g., Microsoft employees, 40 mechanical engineers, a university's computer science department, and five university admins; Agrawal *et al.*, 2007; Hicks *et al.*, 2008; Tanenbaum *et al.*, 2006; Skondric *et al.*, 2020), and none have compared FSDs across today's popular operating systems (which can influence collections' structural properties; Dinneen & Frissen, 2020).

Although data on users' FSDs could help improve everyday tools for retrieving and organising personal information, to our knowledge no general study of users' FSDs has been conducted since 2007, no prior study has examined FSDs among personal collections, and none have compared distributions across Windows, Mac, and Linux.

## METHODOLOGY

To address the above gaps in knowledge we pose the following research questions:

> RQ1. What are the differences in FSDs among today's popular desktop operating systems?
>
> RQ2. What are the FSDs like today in collections used for different purposes, and how do they differ?
>
> RQ3. How much growth in file sizes has taken place since the last major study of FSDs?

*Data Collection and Sample* - 348 remote and anonymous participants downloaded and ran on their desktops and laptops open-source software (Dinneen *et al.* 2016) that collected data about files they indicated they manage. File sizes were measured in bytes using python's `os.stat().st_size`; the collected data are thus comparable to those collected by scanning files or monitoring system traces and disk access (Agrawal *et al.*, 2007; Baker, 1991; Gonçalvez & Jorge, 2003; Roselli, 2000; Vogels, 1999). Care was taken to exclude files not managed by the participant: hidden files and common folders containing operating system files were explicitly ignored. A summary of the sample is in Table 1, and further details are provided in prior work (Dinneen, Julien & Frissen, 2019, p. 4).

| Size | Age | Gender | Operating system | Collection use |
|---|---|---|---|---|
| 348 participants 49 million files | Range: 14-64 Mean: 30 SD: 9.96 | Male: 218 (63%) Female: 123 (35%) Other: 7 (2%) | Mac OS (10.7 – 11): 169 (48%) Windows (XP – 10): 135 (39%) GNU/Linux: 44 (13%) | Work: 166 (48%) Knowl. work: 93 Other: 48 IT: 25 Study: 143 (41%) Personal: 39 (11%) |

**Table 1. Summary of sample**

*Data Analysis* - Collections were divided according to each analysis conducted: first into the popular operating systems (OSes; Windows, Mac OS, and GNU/Linux), and second into the use of the collection (personal matters only, work, or study) according to participants' responses within the data collection interface. Work collections were further divided (using participants' stated occupations) into three groups common in PIM literature and used in past analyses of the same data (Dinneen & Julien, 2019): knowledge workers (e.g., manager, doctor, journalist), IT staff (e.g., programmer, systems administrator), and all others (e.g., tradespeople, retail, artist, unemployed). To facilitate comparison with prior works we generated arithmetic and log-normal averages (median and mean) for each group and visualised their file FSDs using cumulative distribution frequency plots (CDFs), which illustrate how much of a collection files of different sizes occupy (in terms of count, not disk space used). Because the data are extremely skewed, we assessed the statistical significance of differences between groups with the Mann-Whitney U test, a non-parametric equivalent to t-tests. Finally, while we did not aim to exclude outliers, a collection of 2.2 million files was almost entirely comprised of files sized 64 KB, and a collection of one million files was 50% comprised of 4 KB files. We excluded these collections from all analyses, as they dramatically altered the CDFs otherwise.

## RESULTS

Below are highlighted results and implications. Comprehensive CDF plots, full data tables with discrete values for each distribution, and all analysis scripts can be accessed at github.com/jddinneen/fm-results-tables.

### FSD of all data and in each OS

Table 2 summarises the FSDs of the complete data set and those the collections in each OS. Relatively small files are very common in each OS: files below ~5 KB account for 50% of Mac and Linux collections, and files below 8.5 KB account for 50% of files in Windows. However, the CDFs of each OS differ significantly ($p<0.001$) across all measures. Notably, the Mac distribution has more small files than Windows's, while Linux is so skewed (e.g., SD 145 MB, three times that of Windows) that its CDF resembles Mac's below 5 KB (i.e., such files are 50% of both CDFs) despite having more files above 64 KB than Windows.

### FSDs of different collection types

As seen in Figure 1, files larger than 32 MB are relatively rare in collections of all use types, but the FSDs are otherwise significantly different (all pairs $p<0.001$). Personal collections have more large files and are more skewed than work-only collections (log-normal average 15 KB to 1.9 MB), except for 'other' work collections, with which they are largely similar. Among work collections, IT collections are the most different, being the most composed of small files (log-normal average 3.6 KB to 126 KB). Finally, study collections are between IT and knowledge work collections in most regards.

| Data set / OS | Log-normal median & mean | Arithmetic mean | 50% occupied by (< mean) |
|---|---|---|---|
| whole data set | 9.0 KB, 730 KB | 1.5 MB | < 5.4 KB |
| Mac OS | 8.0 KB, 533 KB | 1.4 MB | < 4.9 KB |
| Windows | 11.5 KB, 1.0 MB | 1.7 MB | < 8.3 KB |
| GNU/Linux | 10.8 KB, 1.7 MB | 2.2 MB | < 4.8 KB |

**Table 2. Summary of FSDs of the total data set and within each OS**

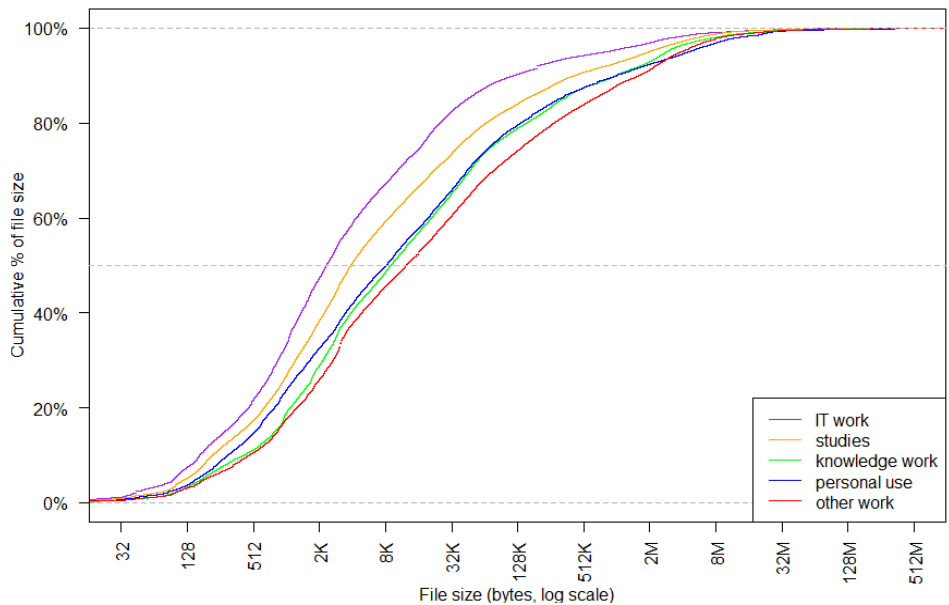

**Figure 1. CDFs of file sizes in collections used for different purposes**

## DISCUSSION

RQ1. *FSDs differ slightly but significantly between today's operating systems.* That larger differences were not seen is perhaps surprising given the influence of the OS on many other properties of the file collection (Dinneen & Frissen, 2020). While the small differences likely do not warrant notable changes in the features of file manager interfaces, they are nonetheless important for the performance of hard drive firmware, media players, and file transfer operations. In particular, performance might be improved for users in all three OSes (and cross-platform scenarios) by increasing the file system block size to reflect the discrete values observed – Windows in particular may benefit from a block size of 8 KB. As the frequency of mobile FM increases (Bergman & Yanai, 2018), it may also be beneficial to examine FSDs across mobile contexts and operating systems (i.e., Android and iOS devices).

RQ2. *FSDs differ between collection uses: personal collections contain more large files than most work collections, and IT collections contain more small files.* These differences are consistent with a recent study of the *types* of files seen in the same data (Dinneen & Julien, 2019): personal collections and 'other' work collections contain fewer documents but more (large) media than other collections, and IT collections contain (small) plain text files for system administration and software development. The similarity of study collections to both IT and knowledge work collections may reflect that studying today resembles knowledge work but also commonly entails IT or programming tasks. Regardless of the cause, we suggest designers of file managers utilise this information, for example by prioritising files in the ranges given above when performing desktop search (i.e., use the file metadata to rank the results) or recommending files to users (e.g., favour files in the typical range for that occupational setting or for that user); whereas the largest files likely stand out in users' memory and are thus easier to find through folder navigation, results that place small files (<1 MB) *higher* would increase the chances that the user sees the file they want since, *ceteris paribus*, the user likely wants a file in that size range. We also suggest future studies further explore the connection between FSDs and file *type* distributions, perhaps *especially* as the merging of personal and work collections presumably increased when people were working from home during the COVID-19 pandemic. It may be useful, for example, to suggest retrieving or deleting files that are exceptionally small or large *for their type*.

RQ3. *File sizes have grown more than ten-fold since the mid-2000s, but most files are still under 8 MB in size.* When comparing our knowledge workers' FSD with those of past results, seen in Figure 2, there appears to be an increase in the presence of larger files from the mid-2000s (Agrawal *et al.*, 2007; Tanenbaum *et al.*, 2006) to today, but without many more *very large* files: whereas such collections were previously mostly (>90%) composed of files smaller than 32 KB, that value today has grown to 1 MB, and files larger than that are still under 8 MB (<95%). These findings are consistent with the only other recent study (Skondric *et al.*, 2020), of five university administrators' collections, except that study found *very few* files are under 8 KB in size (approx. <5%, rather than approx. < 50%). It is unclear if this difference is attributable to particular job roles (i.e., uni admins in particular) or simply that particular sample. Regardless, the overall growth we have identified may be attributable to increasing storage capacity generally and new and richer media formats more specifically, which increase in resolution/fidelity (and therefore size) over time. Further growth may also be obscured by older files created under the constraints of prior OSes and software (i.e., smaller block sizes and maximum file sizes), but the strength of such effect is unclear. Either way, we recommend that developers – of operating systems and file management tools – update their designs to account for today's FSDs, and that future works investigate the effect of past file size constraints (e.g., using historical OS data with file creation time metadata).

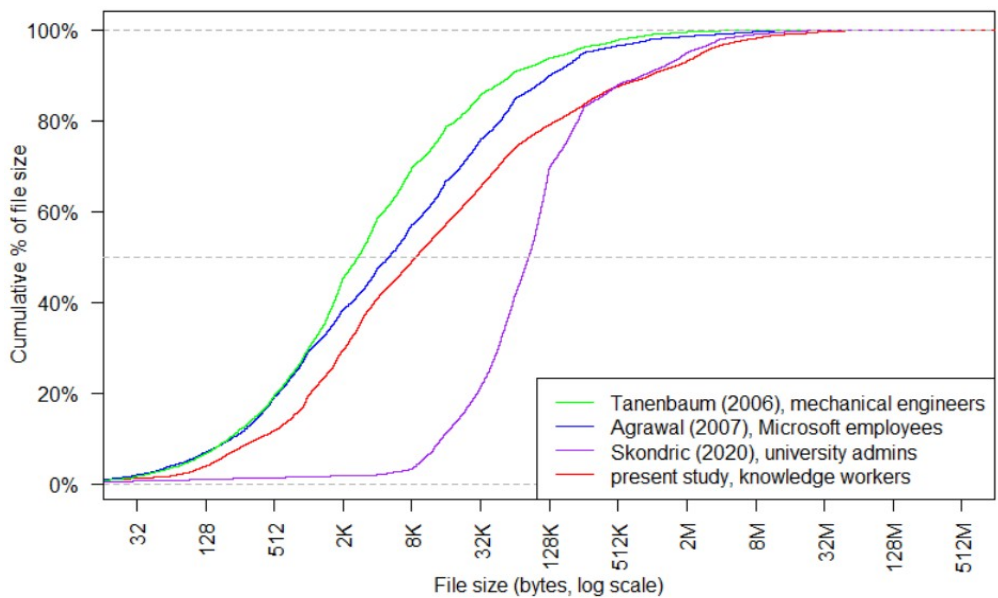

**Figure 2. CDFs of file sizes in knowledge workers' collections from studies past and present**

**CONCLUSION**

In this manuscript we have provided updated data on file size distributions including, for the first time, data about users' personal collections. We have also identified differences among collections types and between operating systems, and traced the changes in knowledge workers' files over time. Finally, we have made suggestions for how our findings could be used to optimise system performance and, importantly, to improve everyday tools for retrieving and organising personal information in people's digital collections. Promising directions for future work, in addition to those discussed above, therefore include designing and testing such improvements, the former of which should be easier with this information about what that test collections should look like if they are to be ecologically valid.

**ACKNOWLEDGEMENTS**

The authors are grateful to the participants for their time and the three anonymous reviewers for their feedback.